\documentclass[10pt,conference,letter]{IEEEtran}

\usepackage{float}

 \usepackage[usenames,dvipsnames]{pstricks}
 \usepackage{epsfig}
\usepackage{epstopdf}
 \usepackage{pst-grad} 
 \usepackage{pst-plot} 
\usepackage{graphics}

\usepackage{algorithm}
\usepackage{algorithmic}
\usepackage{footnote}
\usepackage{cite}
\usepackage{amsmath}
\usepackage{amssymb}
\usepackage{amsfonts}
\usepackage{tikz}

\usepackage[none]{hyphenat}
\begin{document}

\title{Detection of Copy-move Image forgery using SVD and Cuckoo Search Algorithm}

\author{\IEEEauthorblockN{Abhishek Kashyap, Megha Agarwal, Hariom Gupta}
\IEEEauthorblockA{Department of Electronics and Communication Engineering,\\ Jaypee Institute of Information Technology, Noida-201304, Uttar Pradesh, India.\\
Email: abhishek.kashyap@jiit.ac.in, megha.agarwal@jiit.ac.in, hariom.gupta@jiit.ac.in}
\and
}

\markboth{IEEE Transactions On Signal Processing, Vol. XX, No. Y, Month 2013}
{Abhishek and Joshi: Using the style file IEEEtran.sty}

\maketitle
\thispagestyle{plain}\pagestyle{plain}

\begin{abstract}

Copy-move forgery is one of the simple and effective operations to create forged images. Recently, techniques based on singular value decomposition (SVD) are widely used to detect copy-move forgery (CMF). Some approaches based on SVD are most acceptable to detect copy-move forgery but some copy-move forgery detection approaches can not produce satisfactory detection results. Sometimes these approaches may even produce error results. According to our observation, detection result produced using SVD depend highly on those parameters whose values are often determined with experiences. These values are only applicable to a few images, which limit their application. To solve this problem, a novel approach named as copy-move forgery  detection using Cuckoo search algorithm (CMFD-CS) is proposed in this paper. CMFD-CS integrates the CS algorithm into SVD. It utilizes the CS algorithm to generate customized parameter values for images, which are used CMFD under block-based framework.

\end{abstract}


\section{Introduction}

Now a days we are surviving in the era of digital revolution which made it is very easy to a process, access and transfer the information from one place to another. Advanced image processing software makes it extremely vulnerable to tamper a digital image in a daily life. If the image is tempered by single region duplication, the portion of the tampering is difficult to recognize by the naked eyes. In the copy-move forgery, the region is copied from an image and pastes it to another region of the same image.  Therefore we have to check some specific objects in the image. Availability of advanced manipulation tools deteriorates the integrity and reliability of digital image. in our daily life we have to face such type of situation like in media, pictures magazines. Digital photographs are considered by the court all over the world as evidences in different matters but we strongly need to assure the authenticity of the digital images. In following sections we will discuss different type of digital image forgery techniques and their detection process.

\subsection{Methods used for digital –image forgery detection}
Active and passive two approaches are widely used for image forgery detection world –wide \cite{P1_1}.

\subsubsection{Active methods}
   In these methods, we add some authentic information inside the image at the time of capturing this or after the capturing this. Before discriminating to the public it will be processed by watermarking techniques for reliability check.

\subsubsection{Passive methods}
 These methods work on the analysis of the purely digital binary information present in the image there is no need to insert information in the image for authentication.

\subsection{Types of digital image forgeries}
In this section we are discussing about the details of digital image forgeries such as Image splicing, cloning and image re-touching etc. Image re-touching is very less harmful kind of digital image forgery, in which certain features enhanced or reduced from the original image, splicing is created by combination of two or more images for hiding the important information from the original image and cloning is very harmful kind of image forgery problem, which is created by coping one part of the image and paste onto the another part of the same image for hiding some secret information \cite{P1_1}.

Many authors have suggested different methods using different algorithms for digital image forgery detection \cite{P1_3}, \cite{P1_9}--\cite{P1_22}.
But in this paper, we have proposed a new algorithm for detection of copy-move image forgery using SVD and cuckoo search algorithm, this algorithm has many advantages as compared with existing digital image forgery detection methods.

The rest of the paper is organized as follows. A review of image forgery detection is presented in section I. In Section II we present problem formulation for forgery detection. In section III we propose novel  method  based on SVD and CS algorithm for forgery detection. In  section IV we provide experiments and simulation results. In section V we present conclusions and scope of the future work.

\section{Formulation of Problem}

This section analyzes problem in parameter setting after a brief description of the Block based framework.

\begin{figure}
        \begin{center}
        \includegraphics[width=3.6in, height=1.0in]{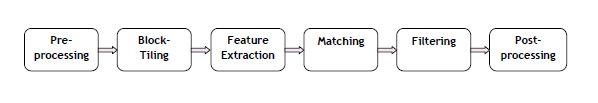}
        \caption{Common flow work of Block-based CMF detection framework}\label{abhi_c}
        \end{center}
\end{figure}

\subsection{The Block-based framework}

CMF detection approaches under the Block-based frame work in a general way that may be divided in to pre-processing. Block tiling, Feature Extraction, Matching, Filtering and Post-Processing as shown in Fig. \ref{abhi_c}.
Pre-processing is to prepare an image for detection such as converting a RGB image in to a grayscale image with standard color space conversion.
Block based method subdivided the image in rectangular regions for every such regions, a feature vector is computed. Similar feature vector are subsequently matched. In block-based method, we consider representative of entire field. They can be graphed in to singular values of reduced –rank.
Feature Extraction is build a descriptor or feature vector, for each key point based on its relationship with the surrounding pixels.
Matching is to determine matched key-points based on feature vector. The regions around the matched key points are probably duplicated regions.
Filtering is to eliminate mismatch key points, which are identified as matched key points during matching, but actually they are not.
Post-processing is to delete duplicated regions or estimate geometric transformation parameters, and so on, when necessary. It depends upon different detection purposes.

\subsection{Problems in parameter values selection}

As detection results depend on the selection of parameter values, an obvious drawback exists in existing CMF detection approach. Normally, these parameter values are determined by experiences or results of test against a number forgery images. However, different research teams choose different values, which are applicable to certain images. The following limitations appear, When they are used to detect a large number of images:
i)	Duplicated regions can not be detected when matched block pairs do not satisfy the matching conditions;
ii)	Detected region are not duplicated ones, if there are too many similar object in  an image and parameter values are chosen inappropriately, some similar regions may be mistakenly regarded as duplicated region, though actually they are native regions in the original images.

\section{DESIGN OF OUR APPROACH}

\begin{figure}
        \begin{center}
        \includegraphics[width=2.3in, height=4.8in]{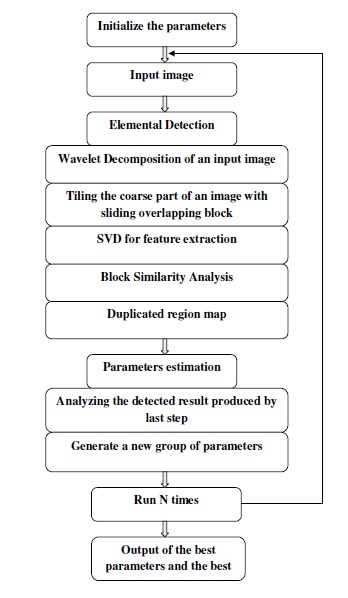}
        \caption{The flow chart of our approach, CMFD-CS}\label{abhi_f}
        \end{center}
\end{figure}

The goal of our approach, CMFD-CS is to automatically generate suitable parameter value for each test image. The flow chart of CSMD-CS is shown in the Fig. \ref{abhi_f}. It includes two components, one of which is elemental detection and the other is parameter estimation. Elemental Detection is derived from the Block based frame work. Its task is to detect CMF images. Parameters estimation is new phenomena, which can generate suitable parameter values for each images. The corresponding image may produce a satisfactory result using these values. The CS algorithm is applied to estimate parameter values. To our knowledge, none of the existing CMF detection approaches use the CS algorithm.

CMFD-CS generates suitable parameter values automatically for each image according to the feature of the image. With these parameter values, Elemental Detection can produce better results.
The first step is to identify the input and the output of the Block-based frame work. The input includes an image and a group of parameters. The output is only the number of matched blocks. Which is used to evaluate whether the result is good. We turn parameter value estimation in to an issue of optimal solution. An evaluation criterion is created to make detection decision. The criterion is formed by the number of matched blocks. When the criterion reaches extreme value optimal solution turn out.

\subsection{Cuckoo Search Algorithm}
For describing Cuckoo-Search Algorithm \cite{PNIA5} in a simple manner we are following three idealized rules: (i) Each cuckoo lays one egg at once, and dump its egg in arbitrarily picked nest; (ii) The best nests with high quality of eggs will continue to the next generations; (iii) The number of accessible host nests are fixed, and the egg laid by a cuckoo is found by the host bird with a probability $p_{a}  \in    [0, 1]$.
For this situation, the host bird can either discard the egg or abandon the nest and build
a totally new nest. For effortlessness, this last presumption can  be approximated by the division $p_{a}$ of the $n$ nest are supplanted by new nests (with new random solutions)\cite{PNIA6}--\cite{PNIA7}.

For a maximization problem, the fitness value or quality of a solution can be simply proportional to the value of the
objective function.

\begin{algorithm}
\caption{Cuckoo Search}
\label{algo:CS}
\begin{algorithmic}
\STATE Objective function $f(z), z = (z_1, ..., z_d)^T$
\STATE Generate initial population of $n$ host nests $z_i (i = 1, 2, ..., n)$
\WHILE {$t < MaxGeneration\  or\ stop-criterion$}
\STATE Get a cuckoo randomly by Levy flights
\STATE Evaluate its quality/fitness $F_i$
\STATE Choose a nest among $n$ (say, $j$) randomly
\IF {$F_i > F_j$ }
\STATE replace $j$ by the new solution
\ENDIF
\STATE Fractions ($P_a$) of worse nests are abandoned and new ones are built
\STATE Keep the best solutions (or nests with quality solutions)
\STATE Rank the solutions and find the current best
\ENDWHILE
\end{algorithmic}
\end{algorithm}

Based on these rules, the fundamental steps of the Cuckoo
Search (CS) can be discussed as the pseudo code.
For a cuckoo $i$, Levy flight is performed for generating new solutions $z^{(t+1)}$:

\begin{equation}\label{}
z_{i}^{(t+1)}= z_{i}^{(t)} + \alpha \oplus Levy(\lambda)
\end{equation}

where $\alpha > 0$ is the step size which should be related to the scales of the problem of interests. In most cases, we can use $\alpha = 1$.

The Levy flight basically gives an irregular walk while
the irregular step length is drawn from a Levy distribution:

\begin{equation}\label{}
  Levy \thicksim u = t^{-\lambda},       (1 \leq \lambda \leq 3)
\end{equation}

which has an infinite variance with an infinite mean. Here the steps basically frame an arbitrary walk process with a power-law step-length distribution with a substantial tail. Some of the new solutions should be generated by Levy walk around the best obtained solutions so far, this will accelerate the local search.
However, a substantial fraction of the new solutions should be generated by far field randomization and whose locations should be sufficiently far from the current best solution, this will ensure that the system would not be trapped in a local optimum. There are some noteworthy contrasts between CS and some other optimization algorithm, randomization is more efficient as the step length is substantially tailed,
and any large step is possible and the number of
parameters to be tuned is less than GA and PSO \cite{PNIA2}. So that it is potentially more generic to adapt to a wider class of optimization problems.

Initially, random or manually generated initialization parameter values are used to detect forged images using CMFD-CS. Then the following two operations are executed N times.
i)	Elemental discussion detects the input images with the detection parameter values and then delivers the detection result to the operation (2).
ii)	According to the result from operation (1) a new group of parameter values to operation (1) and start the next round.
The best detection result is chosen from the operation of N rounds. Then this result and relevant parameter values are output. In our experiment, we set the value of N to 1000

\subsection{The elemental detection}


The proposed elemental detection of digital image forgery method involves the following steps:
1) Wavelet decomposition of given image;
2) Tiling the image with overlapping block;
3) Singular Value decomposition (SVD) of each tiled block;
4) Block similarity analysis;
5) Duplicated regions map creation.

\subsubsection{Wavelet decomposition}

This technique starts with the computation of wavelet transform of the given image, after computing wavelet transform, we have got the high-low bands, the low-high bands and the high-high bands of the image at various scales, then we have processed the coarse part of the image for image imitation recognition.

 we have used Harr wavelet,  $\psi(z)$, which  is orthogonal to the scaling function and it is defined \cite{P2_15} by-
  \begin{equation}\label{}
    \psi \left({\rm z}\right){\rm =}\sum^{\infty }_{{\rm m=-}\infty }{{\left({\rm -}{\rm 1}\right)}^{{\rm m}}{{\rm a}}_{{\rm N-1-m}}\sqrt{{\rm 2}}\emptyset \left({\rm 2z-m}\right)}
  \end{equation}

  wavelet decomposition of the function  $f(x,y)$ in two dimension is defined as \cite{P2_15},
  \begin{equation}\label{}
    f\left(z,y\right)=\sum^{\infty }_{j=-\infty }{\sum^{\infty }_{k=-\infty }{\sum^{\infty }_{l=-\infty }{d_{j,k,l}{\psi }_{j,k}\left(z\right)}}}{\psi }_{j,l}\left(y\right)
  \end{equation}

\vspace{2 mm}
\subsubsection{Tiling the image with overlapping block }
\begin{figure}[H]
        \begin{center}
        \includegraphics[width=8cm,height=3cm]{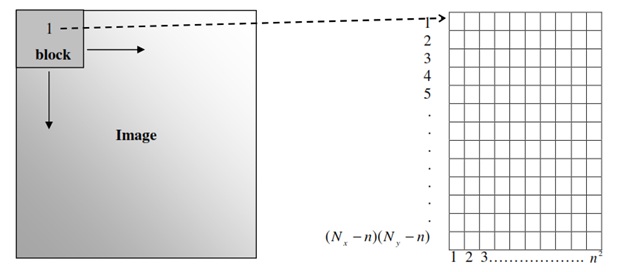}
        \caption{Pixel block scan and array dimensions for the matching algorithm. \cite{P2_12}}\label{abhi3}
        \end{center}
\end{figure}

 The coarse part of the image is being tiled \cite{P1_3} by the block of $(R \times R)$ pixels, which is obtained after wavelet decomposition. This block is horizontally slide from left to right and top to bottom as shown in Fig. \ref{abhi3}. Here we have assumed, the duplicated region size must be larger than block size and the
total overlapping blocks are $(M-R+1)\times(N-R+ 1)$ for digital image size of $(M \times N)$ pixels.

\vspace{2 mm}
\subsubsection{Singular Value decomposition of each overlapping blocks}

With the help of SVD, we can represent each block of digital image \cite{PNIA3},
SVD is a factorization of a real or
complex matrix which is generally used in useful applications of signal processing and statistics \cite{PNIA3_1}. If A is an $m_{1} \times n_{1}$
real or complex matrix, a decomposition of this matrix could
be:

\begin{equation}\label{}
 A = U \Sigma V^{T}
\end{equation}

where $ U \varepsilon (m_{1} \times m_{1})$ and $ V \varepsilon (n_{1} \times n_{1})$ ($V^{T}$ is the transpose of $V$)
are real or complex unitary matrices and $ \Sigma \hspace {1mm}\varepsilon (m_{1} \times n_{1})$ is a
rectangular diagonal matrix in which its diagonal consists of
non-negative real numbers sorted in descending order.
Therefore the diagonal of $\Sigma$ is considered as the singular values
of A.

\begin{equation}\label{}
  S_{v}=diagonal of (\Sigma)
\end{equation}

Correlated variable can be transformed by SVD method in a data set \cite{PNIA3_2} to demonstrate better the different relationship into a set of uncorrelated ones. for the modification of copy-move forgery process, it can model the soft relationship between the column and rows of digital image. So to catch the modified correlation among the digital image pixels, singular values can be applied in copy-move detection technique. A function is needed to give equal emphasis on singular values because copy-move process has various effect on singular values. There for to extract the features of digital images logarithm of inverse power of singular values is introduced. Due to high dependency in an image pixels, the proposed method divide the digital image into sub-block of size $ R \times R $. to evaluate inter block and intra-block correlation.
For feature extraction, we have to calculate the singular value vector ($Sv$) for each sub-blocks $j$:

\begin{equation}\label{}
 Sv_{j}=[\alpha_{1j},\alpha_{2j},...,\alpha_{nj}]
\end{equation}

Then find out the natural logarithm of
inverse of each singular value is calculated and the
results are summed for each sub-blocks $j$:

\begin{equation}\label{}
  SVB_{j}=\sum_{i=1}^{n}\log(\alpha_{ij}^{-1})
\end{equation}

\begin{table*}[ht]
\caption{Optimization parameters (Parameters in Block-based framework)} 
\centering 
\begin{tabular}{c c c c} 
\hline\hline 
Parameters & Meaning & Lower bound & Upper bound \\ [0.5ex] 
\hline 
$R$ & Block size & 4 & 20 \\ 
$D$ & The minimum distance & 10 & 40 \\
$T$ & The parameter useful for rejecting unstable blocks & .001 & .9\\[1ex] 

\hline 
\end{tabular}
\label{table1} 
\end{table*}

\vspace{2 mm}

\subsubsection{Blocks similarity analyses }

In this step, similarity between the sub-blocks is obtained by calculating Euclidean distance, if we found any sub-block has lesser Euclidean distance, at that point we can say they are comparable. This is a necessary condition
but not sufficient.  Also we have to check their neighborhood sub-blocks for finding similarity. If their neighborhood is additionally comparable, at that point there is a high likelihood that they are copied and they should be labeled.

   The similarity measure $S(B_{i}, B_{j})$ \cite{P1_3} between two sub-blocks is defined as:
\begin{equation}\label{}
  S\left(B_i,B_j\right)=\frac{1}{1+\rho \left(B_i,B_j\right)}
\end{equation}

\ Where $\rho $ is Euclidean distance between two sub-blocks:
\begin{equation}\label{}
  \rho\left(B_i,B_j\right)={\left(\sum^{dim}_{k=1}{{\left(B_i\left[k\right]-B_j\left[k\right]\right)}^2}\right)}^{{1}/{2}}
\end{equation}

 If $S(B_{i}, B_{j})$ $\ge $ $T$, (where T is the minimum required similarity), then we have analyzed the neighborhood of $B_{i} $ and $  B_{j}$.
 Threshold ($T$) played a very important role to obtain the degree of reliability between sub-blocks $i$ and $j$, which is used to take a decision for digital image forgery. We have chosen 16 neighboring sub-blocks with a maximum distance of 4 pixels from the analyzed sub-block for analyzing the blocks neighborhood.

    \begin{equation}\label{c}
      S\left(block\left(i+x_r,s+y_r\right),block\left(j+x_r,t+y_r\right)\right)\ge T
    \end{equation}
 \ Where x $\epsilon$ (-4,-3... 4) and y $\epsilon$ (-4,-3\dots  3, 4) and r=1... 16.

  If $S(block(i, s)$, block$(j, t))$ $\ge $ T, but
\begin{equation}\label{d}
  \sqrt{\left({\left(i-j\right)}^2+{\left(s-t\right)}^2\right)}~~\le ~~D
\end{equation}

 Using equation $(\ref{c})$ and $(\ref{d})$, we have obtained the minimum size of forged area. If similarity between sub-blocks is greater than threshold $T$ but distance between them is less than the threshold $D$, then these sub-blocks will not be further examined and will not be assigned as a duplicated region. Threshold D is used to determine the minimum distance between duplicated regions and it plays a important role to provide more precise results for digital image forgery detection.

  Finally We got a result in form of a matrix Q, which will be same size of the coarse part of the input image. An element of this matrix is set to one if the block at this position is duplicated otherwise set to zero.

\subsubsection{Duplicated regions map creation }
  Duplicated regions map is created by the multiplication of each element of $I(x, y)$ by its respective element in $Q(x, y)$

\subsection{The parameter estimation}

The CS algorithm is used to search the adjustable parameters. The CS algorithm is suitable for solving minimization or maximization problems.

\subsubsection{Parameter for elemental detection}
The parameters of the Block-based framework need to be optimized and their boundaries are listed in table \ref{table1}. The reason for the choice of these parameters is that these parameters will make evident effect for the final detection results.

\subsubsection{Evaluation function}

Although metrics for detection approaches are various in different literatures, the core idea are similar: more true matched Blocks $(TMB)$, less mismatched Blocks $(MMB)$ and less missing matched Blocks $(Miss-MB)$. Therefore in the process of building the evaluation function, these factor should be considered.
The Ideal solution is that, the number of the $TMB$ are as large as possible. The large number of $TMB$ are not only conductive to estimating the duplicate region accurately, but also make the detection results more convicting. Therefore, the evaluation function is built as the following:

\begin{equation}\label{}
  P_{match}= \frac{TMB_{t}}{TMB_{t}+ \phi}, \phi=\begin{cases}
               MMB_{t}, MMB_{t}>10\\
               10, MMB_{t}\leq10\\
                         \end{cases}
\end{equation}

Where $TMB_{t}$ and $MMB_{t}$ are not the number of the true matched blocks or mismatched blocks in fact. They are both determining by the affine transform at filtering. The pairs of blocks meeting the affine transform are regarded as true matched key points $TMB_{t}$ and other pair are taken as the mismatched key points $MMB_{t}$ does not include the eliminated pair of matched blocks that the distance between them less than $D$. $\phi$ is the mismatch coefficient and $\phi$ gives a default minimum value for $MMB_{t}$. $P_{match}$ is the probability of real matching. The evaluation criterion of CMFD-CS is $P_{match}$. Parameters Estimation will choose the highest value of $P_{match}$ as the best results.

\section{Experiments and results}

\subsection{Metrics}
The important measures are the number of effectively recognized forged images, $T_{P}$, the number of images
that have been wrongly distinguished as forged, $F_{P}$ and the
falsely missed forged images $F_{N}$ at image level. From these we have computed
the measures $Precision (p)$, and $Recall (r)$. They are defined as:

\begin{equation}\label{}
  p=\frac{T_{P}}{T_{P}+F_{P}}, and r=\frac{T_{P}}{T_{P}+F_{N}}
\end{equation}

Precision indicates the probability that a detected forgery is
a truly forgery, while Recall demonstrates the probability that a forged
image is detected. Recall is also called as true positive rate.

\subsection{Results}

In this section we have discussed the results for the method described in section III.

We have analyzed different images for detection of copy-move forgery from the created database \cite{P1_A5}.

\begin{figure}[]
        \begin{center}
        \includegraphics[width=3.7in, height=1.3in]{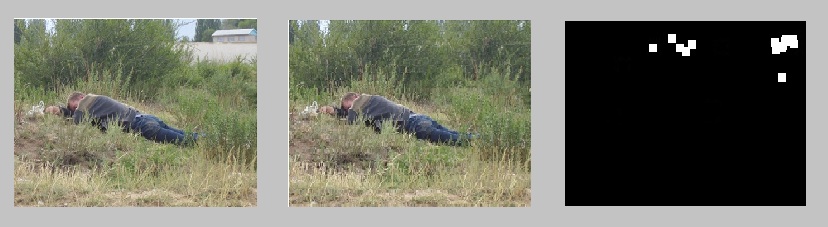}
        \caption{(a) Original image \cite{P1_3} (b) Forged image (c) Duplicated region map}\label{abhi5}
        \end{center}
\end{figure}

\begin{figure}[]
        \begin{center}
        \includegraphics[width=3.7in, height=1.3in]{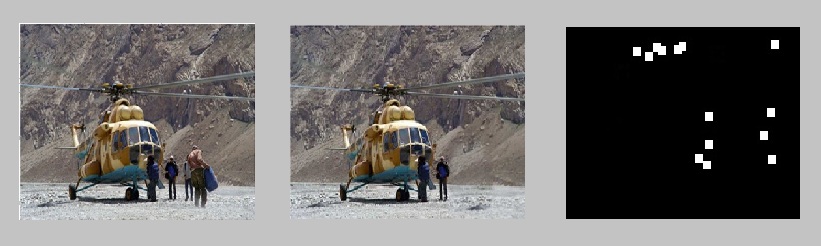}
        \caption{(a) Original image \cite{P1_3} (b) Forged image (c) Duplicated region map}\label{abhi6}
        \end{center}
\end{figure}

\begin{figure}[]
        \begin{center}
        \includegraphics[width=3.7in, height=1.3in]{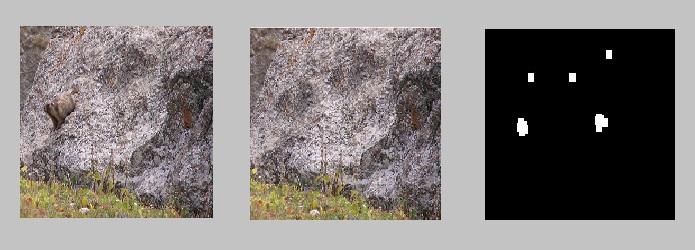}
        \caption{(a) Original image \cite{P1_21} (b) Forged image (c) Duplicated region map}\label{abhi7}
        \end{center}
\end{figure}

\begin{figure}[]
        \begin{center}
        \includegraphics[width=3.7in, height=1.3in]{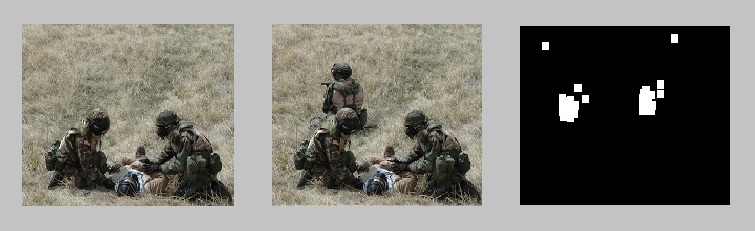}
        \caption{(a) Original image \cite{P1_3} (b) Forged image (c) Duplicated region map}\label{abhi8}
        \end{center}
\end{figure}

\begin{figure*}
\centering
       \includegraphics[width=6.5in, height=2.4in]{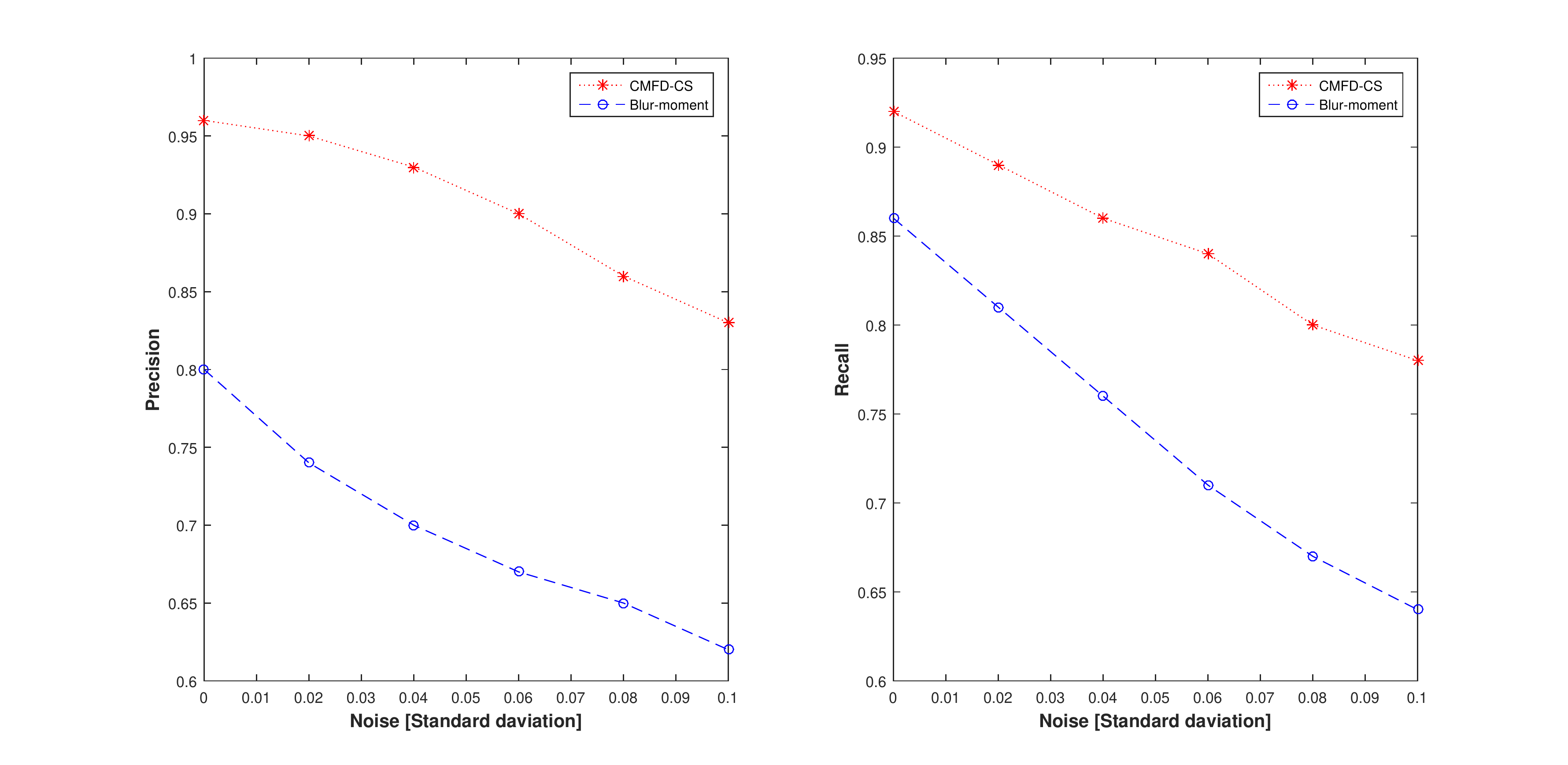}
  \caption{Comparison between CMFD-CS and blur-moment \cite{P1_3} based algorithm, when adding Gaussian Noise.}\label{abhi9}
\end{figure*}

\begin{figure*}
\centering
       \includegraphics[width=6.5in, height=2.4in]{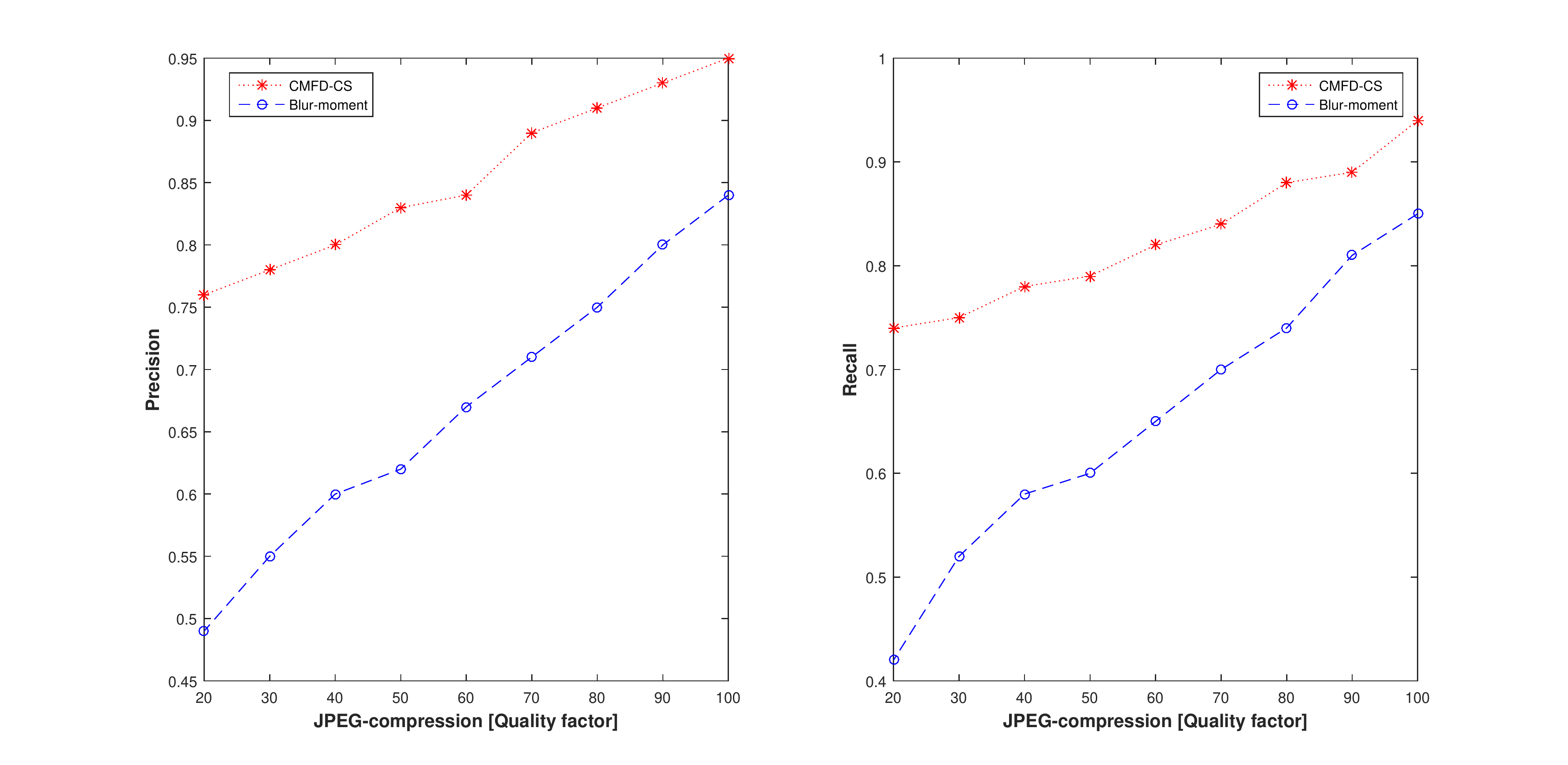}
  \caption{Comparison between CMFD-CS and blur-moment \cite{P1_3} based algorithm, when JPEG Compression}\label{abhi10}
\end{figure*}

\begin{figure*}
\centering
       \includegraphics[width=6.5in, height=2.4in]{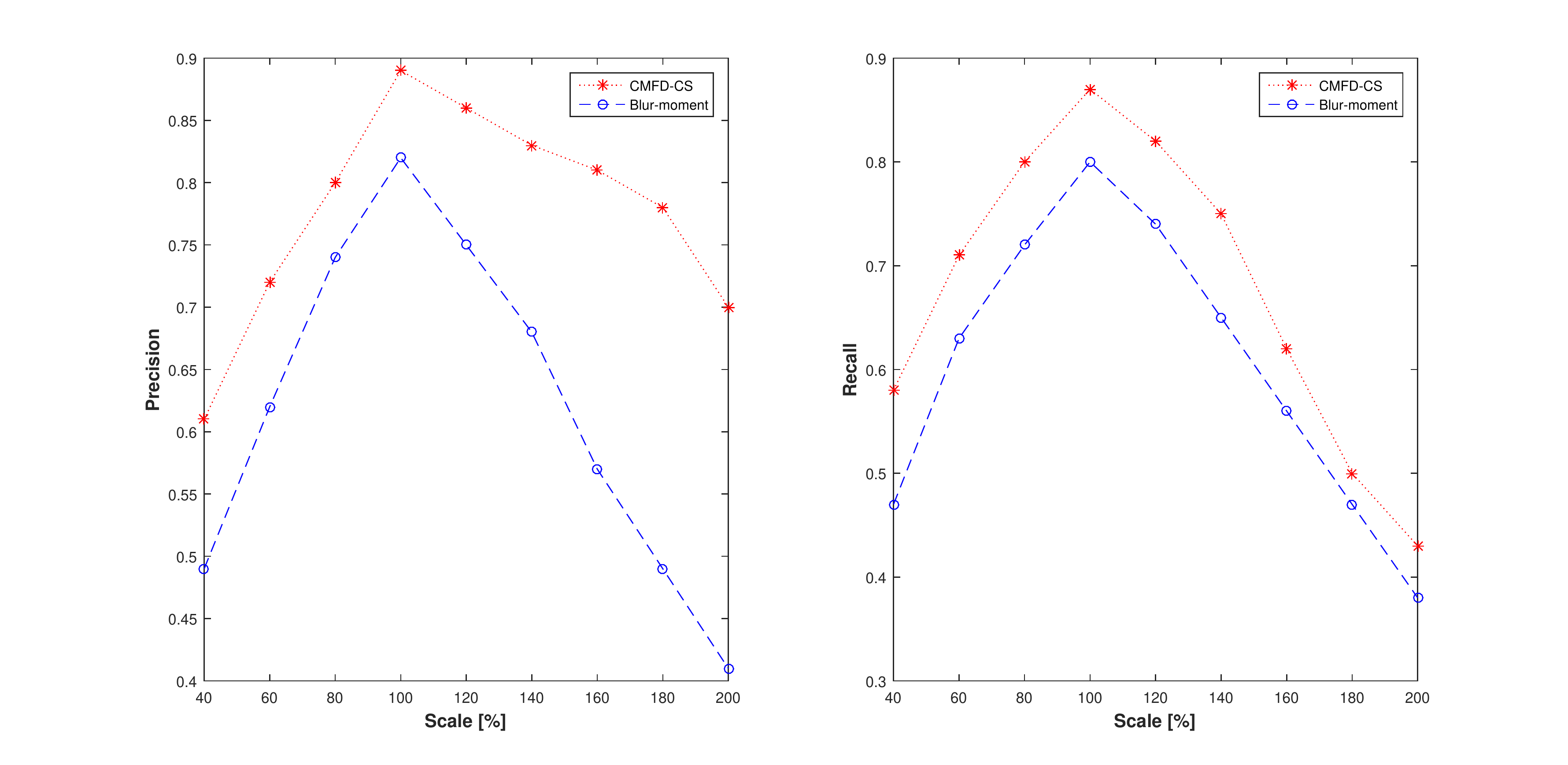}
  \caption{Comparison between CMFD-CS and blur-moment \cite{P1_3} based algorithm, when Scaling}\label{abhi11}
\end{figure*}

Original image, forged image and duplicated image are shown in Fig. \ref{abhi5}, \ref{abhi6}, \ref{abhi7} and \ref{abhi8} respectively. When forged images has processed to our proposed algorithm for detection of copy-move forgery, we have got the possible sign of the tempering.

We test the robustness of CMFD-CS against various attack, which include plain copy-move, Gaussian noise, JPEG compression and scaling, the population size is 50 and maximum number of fitness evaluation is 1500.  The duplicated regions will be translated as follows:

\begin{itemize}
  \item Plain copy-move: Duplicate region is moved to the target location without any additional modification.
  \item Add Gaussian Noise: The images intensities are normalized between 0and 1 and add zero mean Gaussian noise with standard deviations of 0.02, 0.04, 0.06, 0.08 and 0.10 to the duplicate regions.
  \item JPEG compression: JPEG compression is common global disturbance. The quality factor varied between 100 and 20 in steps of 10 degree.
  \item Scaling: The duplicate regions are rescaled by 40\%, 60\%, 80\%, 100\%, 120\%, 140\%, 160\%, 160\% and 200\%.
\end{itemize}

Our proposed method has Precision and Recall 0.96 and 0.92 respectively for plain copy-move forgery, when no noise has been added. This was obtained, when we have performed a blind experiment containing an unkown mixture of 2500 authentic and forged images. Precision and Recall has been reduced, when we add zero mean Gaussian noise with standard deviations of 0.02, 0.04, 0.06, 0.08 and 0.10 to the duplicate regions, which is shown in Fig. \ref{abhi9}.
Same as, Precision and Recall 0.95 and 0.93 respectively for plain copy-move forgery, when no JPEG compression has been performed, Precision and Recall has been reduced, when quality factor varied between 100 and 20 in steps of 10 degree. which is shown in Fig. \ref{abhi10} and Precision and Recall 0.89 and 0.87 respectively for plain copy-move forgery, when no scaling has been performed, it's values reduce to lower value, when duplicate regions has been rescaled by 40\%, 60\%, 80\%, 120\%, 140\%, 160\%, 160\% and 200\%, which is shown in Fig. \ref{abhi11}.



\section{Conclusion and scope for future work}

In this paper, we propose a novel approach CMFD-CS to detect CMF in digital images. Comparing with existing work, this paper makes three contributions.
(i)	It puts forward the concept of applying the CS algorithm to CMF detection.
(ii)	It integrates the CS algorithm into the Block-based Framework to perform CMF detection.
(iii)	It divides rules to automatically determine customized parameter values for given images that are to be detected.
Experimental results show that CMFD-CS can automatically generate customized parameter values for images. Which are independent of neither experiences nor experiments. CMFD-CS can achieve much better results than existing techniques. It can identifies matched points that its counterpart can not, and it can dramatically increase the number of true matched blocks, which make the detection of duplicated region more accurate and more acceptable.
Although CMFD-CS is applicable to most of the CMF images. We find that the approaches under the Block based frame work can not find reliable blocks in uniform texture region or when the duplicate regions are too small.

\nocite{*}
\bibliographystyle{IEEE}

\end{document}